\documentclass[letterpaper]{article} %
\usepackage{aaai2026} 
\usepackage{times}  %
\usepackage{helvet}  %
\usepackage{courier}  %
\usepackage[hyphens]{url}  %
\usepackage{graphicx} %
\usepackage{natbib}  %
\usepackage{caption} %
\usepackage{algorithm}
\usepackage{algorithmic}
\usepackage{lipsum}
\usepackage{xcolor}

\usepackage{newfloat}
\usepackage{listings}
\usepackage[most]{tcolorbox}
\usepackage{booktabs}
\usepackage{array} 
\usepackage{longtable}

\usepackage{tikz}
\usetikzlibrary{shapes.geometric, arrows.meta, positioning}

\DeclareCaptionStyle{ruled}{labelfont=normalfont,labelsep=colon,strut=off} %
\floatstyle{ruled}
\newfloat{listing}{tb}{lst}{}
\floatname{listing}{Listing}
\definecolor{SoftGreen}{HTML}{C9E7C9}

\title{A Justice Lens on Fairness and Ethics Courses in Computing Education: LLM-Assisted Multi-Perspective and Thematic Evaluation
}

\author{
    Kenya S. Andrews\equalcontrib\textsuperscript{\rm 1},
    Deborah Dormah Kanubala\equalcontrib\textsuperscript{\rm 2},
    Kehinde Aruleba\textsuperscript{\rm 3},
     Francisco Enrique Vicente Castro\textsuperscript{\rm 4},
     Renata A Revelo\textsuperscript{\rm 5}
}
\affiliations{
    \textsuperscript{\rm 1}Brown University,
    \textsuperscript{\rm 2}Saarland University,
    \textsuperscript{\rm 3}University of Leicester,
\textsuperscript{\rm 4}New York University,
\textsuperscript{\rm 5}University of Illinois \\

}

\usepackage{bibentry}

\begin{document}

\maketitle

\begin{abstract}
Course syllabi set the tone and expectations for courses, shaping the learning experience for both students and instructors. In computing courses, especially those addressing fairness and ethics in artificial intelligence (AI), machine learning (ML), and algorithmic design it is imperative that we understand how approaches to navigating barriers to fair outcomes are being addressed.These expectations should be inclusive, transparent, and grounded in promoting critical thinking. Syllabus analysis offers a way to evaluate the coverage, depth, practices, and expectations within a course. Manual syllabus evaluation, however, is time-consuming and prone to inconsistency. To address this, we developed a \emph{justice-oriented scoring rubric} and asked a large language model (LLM) to review syllabi through a multi-perspective role simulation. 
Using this rubric, we evaluated 24 syllabi from four perspectives: instructor, departmental chair, institutional reviewer, and external evaluator. We also prompted the LLM to identify thematic trends across the courses. 
Findings show that multi-perspective evaluation aids us in noting nuanced, role-specific priorities, leveraging them to fill hidden gaps in curricula design of AI/ML and related computing courses focused on fairness and ethics. 
These insights offer concrete directions for improving the design and delivery of fairness, ethics, and justice content in such courses.
\end{abstract}

\section{Introduction}
\label{sec:introduction}
Artificial Intelligence (AI) education is evolving rapidly as instructors, institutions, and policymakers explore how best to prepare students for a world where AI is increasingly applied~\cite{stolpe2024artificial, milberg2025whyAI}. As AI systems are included in the decision making process in critical contexts such as healthcare~\cite{chustecki2024benefits}, finance~\cite{bhat2024application}, and criminal justice~\cite{taylor2023justice}, among others, the ways we equip learners to critically engage with issues of fairness, bias, and justice in understanding, designing, and deploying AI systems should be highly intentional~\cite{grosz2018embedded, baumer2022integrating, folorunso2025higher}. Importantly, this imperative is not limited to students pursuing AI-related careers. AI literacy, including the ability to understand its broader societal impacts, is becoming essential across disciplines, from law and public policy to education and the humanities~\cite{xiao2025ai, tadimalla2024ai}.

In higher education, course syllabi play a foundational role in setting expectations, framing course values, and communicating institutional priorities~\cite{gannon2018create, gauthier2025syllabus, wiese2024department}. In AI education, syllabi are often the first, and sometimes the only, document that signals whether justice-oriented perspectives are embedded in the learning environment. This includes whether harms of traditional approaches are named, diverse perspectives are incorporated, and steps toward fairness are explicitly discussed. While prior work has examined ethics in computing syllabi~\cite{mccormackinvestigating, dobesh2023towards, hooper2024values, tong2025we}, few have analyzed fairness and ethics in AI, machine learning (ML) and algorithmic design across institutions and course levels, considered them from multiple evaluator perspectives, or employed a justice-oriented lens for reviewing them. Evaluations of syllabi are not neutral, but are often done in silo. Evaluators like internal instructors, departmental chairs, institutional reviewers, external evaluators, and others each prioritize different elements of course design due to holding different values~\cite{velthuis2021educators, alexandre2024systematic} and naturally hold their own biases. These varying role-based perspectives can be valuable, but should not obscure critical opportunity gaps in teaching justice-oriented content in AI/ML fairness and ethics courses. %
Moreover, manual review of these syllabi at scale is labour-intensive and time-consuming, whilst causing limiting opportunities for systematic comparison. 

In this paper, we present a \emph{multi-perspective analysis} of undergraduate and graduate AI/ML syllabi from a range of institutions located in the United States (U.S.), including Predominantly White Institutions (PWIs), Minority Serving Institutions (MSIs), and Historically Black Colleges and Universities (HBCUs), using a \emph{justice-oriented scoring rubric}. We denote \emph{justice} throughout this work as the idea of encouraging outcomes of a society (i.e., thought) such that its members are given what they need based on the particular barriers that infringe on them being an active, free, participating member of that society. Justice furthers the concept of equity in that you do not only receive what helps you move around barriers, but what should be allotted to you based on those barriers to help you remove them~\cite{mobilizegreen2018equityjustice}. Justice also involves understanding what a community needs rather than meeting a perceived need~\cite{benjamin2022viral}. Thus, we apply a lens of justice to assess the aims and tone of courses which are designed to encourage fair and ethical AI/ML and related areas of computing.

While our study focuses on courses explicitly covering fairness and ethics in AI, we also include related computing courses (e.g., ethical computing, trustworthy ML, governance) to examine how AI-related or AI-influenced concepts are discussed in conjunction with issues of fairness and ethics. We note 64\% of the courses we study are completely situated in AI, while 75\% explicitly cover AI topics, and others include data, privacy, history, etc. We prompt a large language model (LLM) to simulate multiple evaluator roles, enabling a scalable yet structured way to surface both consensus and divergence in evaluations. We also prompt the LLM to find thematic trends across the courses, understanding the consistencies and variances that occur across institutional types and course levels. Our framework focuses on two justice-related dimensions: i) inclusive and diverse learning practices (e.g., authorship diversity, accessibility, participatory design) and ii) substantive fairness education (e.g., contextualizing fairness metrics, naming harms, addressing historical inequities). Our analysis is guided by the following research questions:
\begin{enumerate}
    \item To what extent do LLM-simulated evaluators agree in their assessment of AI/ML and related computing syllabi related to justice aspects, and where do their judgments diverge?
    \item  What justice-oriented practices can be consistently identified across syllabi of courses meant to explicate ideas of fairness and ethics in AI/ML and related areas of computing, and which remain underrepresented? 

    \item How are fairness, bias, and justice positioned in fairness and ethics in AI/ML and related computing courses, and how do they explicate these ideas?

\end{enumerate}
\noindent
We introduce a structured \emph{justice-based scoring rubric} for analyzing fairness and ethics in AI/ML and algorithmic design courses. Paired with scalable, multi-perspective LLM analysis that complements human evaluation, our approach surfaces both visible and hidden justice-related opportunities. This integration reveals thematic and scored areas of improvement in curriculum design, enabling institutions to benchmark practices and advance more inclusive, justice-centered AI and computing education.

\section{Related Works}
\label{sec:related_works}
\paragraph{Justice-Oriented AI Education.}
Calls to integrate fairness, bias, and justice into AI and other computing curricula have grown as the societal impacts of algorithmic systems become more visible~\cite{grosz2018embedded, baumer2022integrating, folorunso2025higher}. 
~\citet{burton2017ethical} argues that AI curricula should go beyond training practitioners to design systems, and instead equip students to critically assess their implications in domains such as healthcare, finance, and criminal justice. Prior work has examined the inclusion of ethics-related content in computing and data science courses, identifying gaps in coverage and calls for deeper integration into core technical training~\cite{mccormackinvestigating, dobesh2023towards, hooper2024values, tong2025we, kwao2023ai}. %
While these studies present important trends, they often treat fairness and ethics as isolated or supplemental content, without interrogating how justice is framed or resisted in course design. Moreover, they rarely examine how justice-oriented elements are positioned throughout courses, vary across institutional contexts, or how they are understood from multiple evaluator perspectives. 

\paragraph{Syllabi as a Reflection of Educational Priorities.}
Syllabi serve as artifacts that communicate institutional values, present an outline of the course structure, and convey the instructor’s pedagogical stance~\cite{gauthier2025syllabus, wiese2024department}. They can signal commitments to inclusion through, for example, author diversity in readings, accessibility provisions, participatory learning activities, and the framing of course objectives. In courses focused on fairness and ethics, syllabi may be the primary site where justice-oriented content is articulated. 
Analyses of syllabi have demonstrated their usefulness for identifying patterns in course design, yet few studies focus on how justice-oriented practices are operationalized in fairness and ethics related AI/ML and related computing courses syllabi, espcially at scale. Moreover, syllabi mediate expectations between instructors, students, departments, and accreditation bodies. As such, their language, and what is omitted, can reveal tensions between institutional aspirations and justice-in-practice for AI/ML and related computing courses.

\paragraph{Evaluator Role and Perspective in Curriculum Assessment.}
Evaluations of syllabi are not neutral. Instructors, departmental chairs, institutional reviewers, and external evaluators may each prioritize different aspects of course design, shaped by their roles, contexts, and institutional priorities~\cite{velthuis2021educators}. Such differences may produce inconsistent judgments about justice-related elements in fairness and ethics in AI/ML and related computing courses courses. Moreover, searches for these imperatives at varying levels of granularity may render vital justice-oriented components visible to some evaluators but not others. For example, institutional reviewers may focus on compliance with formal templates, while external evaluators may attend more to clarity of learning goals, accreditation requirements, or the inclusion of bias mitigation strategies. \citet{alexandre2024systematic} manually evaluated multiple interdisciplinary syllabi using an evaluation rubric and multiple reviewers to ensure consistency and depth in program alignment. Thus, we leverage the idea of multi-perspective analysis to alleviate struggles in assessing syllabi focused in expressing ideas of fairness and ethics as justice-related content.

\paragraph{Scaling Syllabus Evaluation Methods.}
Manual syllabus review can provide rich insights, but it is resource-intensive and time-consuming, making it difficult to apply consistently across large, diverse course sets~\cite{alexandre2024systematic, fischer2022salient}. Course observations have also served as an avenue for evaluation and offer rich insight. However, studies have found that examining syllabi can be a more feasible and scalable proxy for assessing course quality across many courses~\cite{fischer2022salient}. Some studies have used rubrics and multiple human reviewers to increase reliability, but these approaches still face scalability limits~\cite{villa2020inter, messer2025consistent}. Recent advances in natural language processing, including LLMs, offer opportunities to automate aspects of syllabus analysis while preserving structured evaluation frameworks~\cite{xu2025course, gan2023large}. 
However, little research has examined their use for simulating multiple evaluator perspectives in justice-oriented curriculum assessment. 

\smallskip
\noindent
\emph{In contrast to prior studies}, our work combines a structured, justice-oriented rubric with LLM-simulated evaluator roles to systematically benchmark AI/ML and related computing courses syllabi across institution types and course levels. We also search for justice-oriented thematic trends across fairness and ethics-related courses in AI/ML and related computing courses. This combined approach enables scalable evaluation and reveals justice-related strengths and gaps that might otherwise remain hidden.

\begin{table*}[!ht]
\centering
\resizebox{\textwidth}{!}{
\begin{tabular}{p{6cm} p{2.5cm} p{3.3cm} p{3cm} p{4cm}}
\toprule
\textbf{Criteria} & \textbf{0 – Not Evident} & \textbf{1 – Partially Evident} & \textbf{2 – Evident} & \textbf{3 – Fully Evident} \\
\midrule
\textbf{Diversity of Assessment Style} Does the course use traditional and non-traditional assessment methods 
(e.g., Traditional: exams, quizzes, essays | Non-Traditional: peer assessment, journal, debate)? & Only traditional assessments used & Majorly traditional assessments used, few non-traditional & Majorly non-traditional assessments used, few traditional & Diverse range of non-traditional and traditional assessments \\
\midrule
\textbf{Expected Learning Outcomes (specific, measurable aims)} Are the Learning Outcomes connected to algorithmic fairness? & No clear connection with algorithmic fairness & Vague or general outcomes connected to algorithmic fairness & Clear outcomes connected to algorithmic fairness & Detailed, competency-specific outcomes connected to algorithmic fairness \\
\bottomrule
\end{tabular}
}
\caption{Example of items from our justice-oriented scoring rubric for syllabus evaluation.}
\label{tab:exrubric1}
\end{table*}

\section{Data Collection and Preprocessing} \label{sec:data}

We sourced syllabi focused on fairness and ethics in AI/ML and related areas of computing through crowdsourcing at a conference focused on computational methods for increasing access and equity for marginalized people; conducting online keyword searches with variances of the phrases: ``fairness in AI course syllabus", ``ethical computing course syllabus", and ``ML fairness course syllabus"; and online crowdsourcing. The majority of our sample came from publicly accessible online repositories and all were affiliated with U.S.-based universities and colleges. We transferred the syllabi to JSON format, extracting relevant metadata such as institution name, institution type (PWI, HSI, HBCU), Carnegie ranking, course name, department, course level (Undergraduate/Graduate), instructor(s), and course term. To ensure sufficient quality of the syllabi, we manually evaluated them to ensure that they minimally contained the course name, institution name, course description/objectives, grading policy, required resources, expected course activities, and required prerequisites. For other metadata not included, we searched the institutional website of each syllabus. 

We collected a total of 24 syllabi from 16 institutions, including three R2 universities, two Hispanic-Serving Institutions (HSIs), and one Historically Black College or University (HBCU). There were 18 graduate-level course syllabi and 6 undergraduate-level course syllabi. Though we focused on classes that were based in AI/ML and related computing courses, only 16 of the syllabi came from to Computer Science Departments---others include Public Policy, Law, Informatics, Biostatistics, and Data Science. While our sample size (n = 24) is modest, it reflects the accessibility constraints of syllabi in this domain (due to paywalls and institutional restricted access). Crowdsourcing yielded very few returns to our ask. However, thematic analysis benefits from smaller, information-rich samples that allow for in-depth examination of content, allowing us to focus strategically on institutional nuances. Our focused approach enables a nuanced thematic analysis and sets the stage for future large-scale studies encompassing a broader range of syllabi.

\section{Methodology}
\label{sec:methodology}
We utilized OpenAI’s GPT-4o-mini
(paid tier 1) to analyze the syllabi we collected along two pathways: (1) using personas of experts who may regularly assess the usability of syllabi to students, practicality of courses' goals, and alignment of courses' objectives with a department or institution and (2) conducting thematic analysis to understand trends that may venture across courses and institutions. The multi-perspective persona-based analysis allowed us explore how justice-oriented content may be differently perceived depending on the evaluator's role (RQ1, RQ2, RQ3), while the thematic approach allowed us to identify common trends in inclusion, omission, and framing across fairness and ethics syllabi (RQ2, RQ3).

We selected GPT-4o-mini because it demonstrates strong performance in text comprehension and zero-/few-shot reasoning, including an 82\% score on MMLU surpassing comparable models like Gemini Flash while maintaining efficiency as a smaller, cost-effective variant of GPT-4o ~\citep{openai2024gpt4omini, reuters2024gpt4omini}. For parameter choices, we fine-tuned the following GPT-4o-mini model’s parameters: top\_p = 1.0, max\_tokens = 2000, and temperature = 0.1. The \emph{top\_p} parameter controls the range of possible outputs based on the model's generated probabilities of output choices. We chose the most liberal top\_p value to allow for full exploration of response. The \emph{max\_tokens} parameter moderates the length of response from the model. We allow for a lengthy response as we requested responses to be in report format and wanted to capture as many possible trends in the syllabi as possible for the thematic analysis, while not limiting explanation of scoring for the multi-perspective persona-based analysis. The \emph{temperature} parameter regulates the creativity in the model response. We chose a low temperature to reduce hallucinations and emphasize the reporting structure, since we are focusing on specific counts and theme appearances within the syllabi.

\subsection{Justice-Oriented Scoring Rubric}
We designed a justice-oriented scoring rubric (Appendix Table~\ref{tab:rubric1} and \ref{tab:rubric2}) to assess both (i) inclusive and diverse learning practices (e.g., authorship diversity, accessibility, participatory activities) and (ii) substantive fairness education (e.g., contextualizing fairness metrics, naming harms, addressing historical inequities). The rubric comprised of $20$ criteria, with scores $0–3$ (Not Evident → Fully Evident) and operational definitions to promote scoring consistency. Existing rubrics rarely foreground justice as a primary evaluative lens; hence, the importance of our innovative framework.

Our justice-oriented scoring rubric was developed through an \emph{iterative, collaborative process} involving weekly working group sessions over the course of $10$ weeks. 
Our rubric crafting team, who developed the rubric, comprised of seven evaluators who conducted four rounds of reviewing, testing, and editing. During this process, we manually evaluated some syllabi to tailor on the rubric. The team included researchers and educators with backgrounds in various STEM fields in Computer Science (AI, ML, Fairness), Engineering (Computer, Electrical, Mechanical), AI ethics, ML justice, computing education, engineering education, decolonial curriculum design, and critical pedagogy. The team is also composed of members from marginalized backgrounds and identities and included members non-native to the U.S. Example rubric items are shown in Table \ref{tab:exrubric1}(full rubric is in the Appendix Table \ref{tab:rubric1} and Table \ref{tab:rubric2}). The developed rubric serves as the foundation for both the persona prompts and the thematic coding, enabling us to assess the framing and depth of justice practices in the fairness and ethics courses and not just its presence.

\subsection{LLM Syllabi Evaluations}
We prompted GPT-4o-mini (described in Section \ref{sec:methodology}) to assist in our multi-perspective and thematic analyses of the collected syllabi.
\subsubsection{Multi-Perspective Persona Analysis}
To capture role evaluations, we prompted GPT-4o-mini to simulate four evaluator perspectives---\emph{Instructor}: an internal course designer prioritizing pedagogical clarity and feasibility; \emph{Department Chair}: a senior academic leader focused on program alignment and accreditation; \emph{Internal Evaluator}: an institutional evaluator focused on compliance and quality assurance; and \emph{External Evaluator}: a neutral reviewer assessing clarity and societal impact without institutional context.

We developed our prompts through multiple rounds of critical iteration and reflection. Within each round, we discussed points of improvement and practicality to refine the prompts. This approach ensures that they capture both implicit biases and formal responsibilities that different evaluators might bring to a syllabus review. As a result, we ensured that each evaluator received a tailored prompt specifying their priorities, constraints, and evaluation style. Syllabi text was provided in full through the JSON file we describe in Section \ref{sec:data} to the model. The model was then instructed to produce both criterion-level scores and brief justifications. In Figure \ref{fig:LLM_DC}, we see the prompting used to simulate the Departmental Chair (all prompts listed in the Appendix).  
\begin{figure}[!ht]
    \centering
\includegraphics[width=0.4\textwidth]{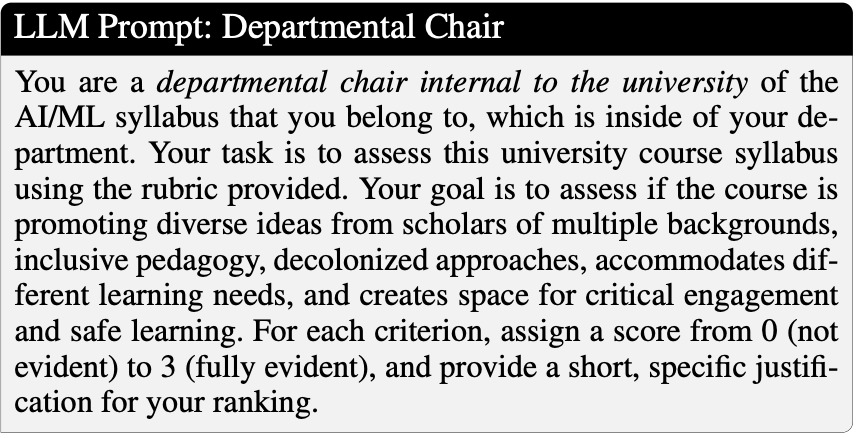}
    \caption{LLM Thematic Prompt: Quantitative Findings}
    \label{fig:LLM_DC}
\end{figure}

\subsubsection{Thematic Analysis}
We conducted a thematic analysis to extract the main themes of our set of syllabi. Beyond identifying shared course topics, our analysis investigated how syllabi stimulated critical engagement and technical practices in relation to fairness and ethics in AI/ML and related areas of computing. We observe this through finding whose voice is centered throughout the course (e.g., if resources written by authors of marginalized backgrounds are commonly used, if community participation is encouraged in algorithmic design, etc), and if diverse ways of learning are used to help digest the material (e.g., game play, small-group discussion, project-based learning). We verified the counts by the LLM with manual analysis for reporting to prevent inaccuracies and hallucinations. This approach underscores the importance of using LLMs as assistive tools rather than replacements for human effort.

Following this, we first explored if there were any general trends and statistically significant findings we could gather from our set of syllabi with a simple prompt as shown in \textbf{LLM Thematic Prompt: Qualitative Finding} in the Appendix \ref{box:themegen}. The LLM struggled here with being able to accurately quantify simple counts, despite us having a strict temperature score---hallucinating more syllabi than actually existed in the dataset.

\begin{figure}[!ht]
    \centering
\includegraphics[width=0.4\textwidth]{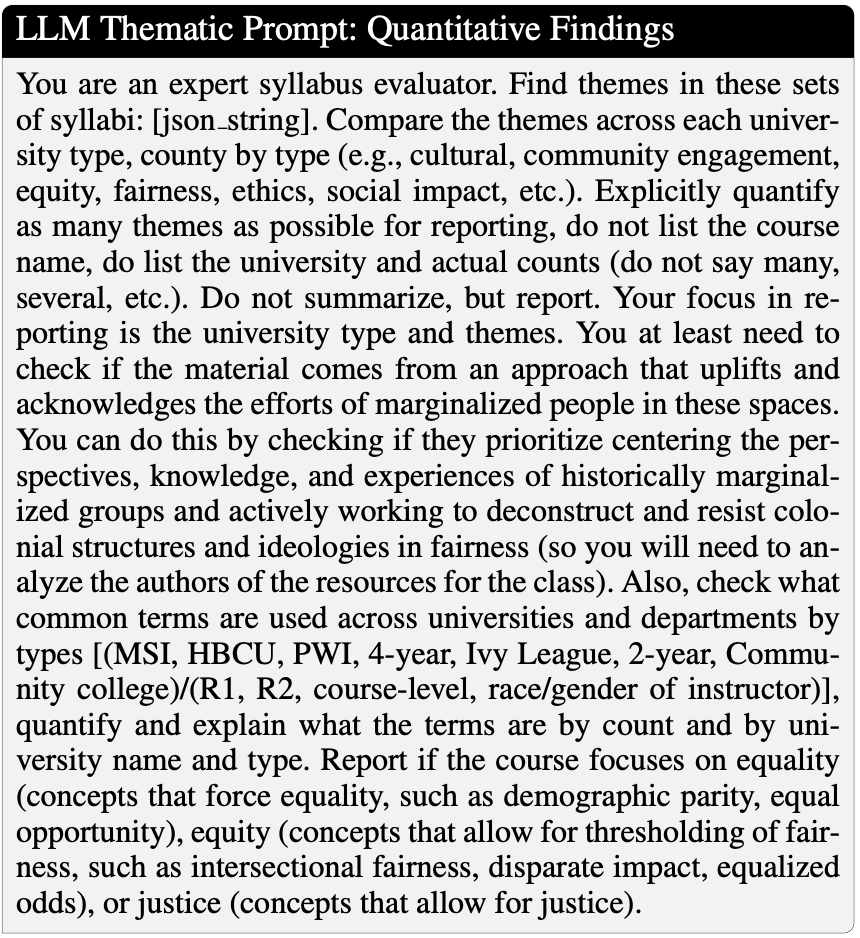}
    \caption{LLM Thematic Prompt: Quantitative Findings}
    \label{fig:LLM_Theme2}
\end{figure}

We then investigated how we can draw out more specific quantities within themes across courses. Our prompting here needed to be highly specific, as shown in Figure \ref{fig:LLM_Theme2}. With looser terms, we noticed over generalization with a lack of quantification (i.e., instead of quantities, terms like ``many", ``several", ``few" were returned) and additional information outside of what was relevant to our analysis, such as introductions to sections of text and two sets of summaries. 

\section{Results \& Analysis}
\label{sec:results_and_analysis}

\subsection{Multi-Perspective Persona Output Analysis}
\subsubsection{RQ1: To what extent do LLM-simulated evaluators agree in their assessment of AI/ML and related computing syllabi related to justice aspects, and where do their judgments diverge?}
\begin{figure}[!ht]
    \centering
\includegraphics[width=0.35\textwidth]{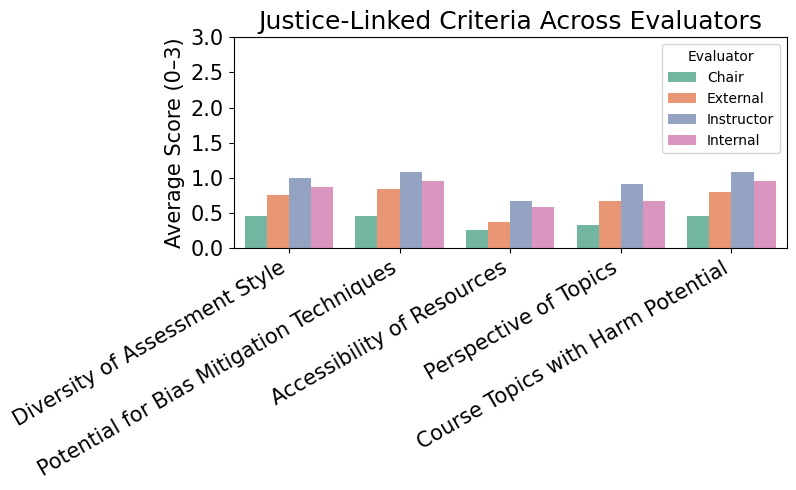}
    \caption{Average scores for justice-linked criteria across evaluators.}
    \label{fig:equity_eval}
\end{figure}

Figure~\ref{fig:equity_eval} compares average scores for fairness-relevant syllabus elements, including \emph{Diversity of Assessment Style}, \emph{Potential for Bias Mitigation Techniques}, and \emph{Accessibility of Resources}. Instructors and Internal Evaluators show strong agreement, consistently rating these criteria higher and similarly. In contrast, Department Chairs diverge by giving markedly lower scores, particularly on \emph{Accessibility of Resources} and \emph{Perspective of Topics}. External Evaluators generally fall between the two groups, aligning with Department Chairs most closely in their assessment of \emph{Accessibility of Resources}. 
The divergences show that multiple perspectives help identify the breadth of possibilities for a course. However, using a rubric in such analyses addresses issues of granularity when assessing justice-oriented practices in courses and for course content. Our findings suggest that justice-oriented elements may be more visible to those directly engaged in teaching (i.e., Instructors, Internal Evaluators) and less prioritized at higher administrative levels (Chairs). By clearly highlighting critical gaps, the rubric supports intentional interventions that improve inclusivity in classroom participation, covering topics that address societal harms by AI/ML design, and strengthen the depth of coverage through incorporating diverse voices for courses addressing fairness and ethics in AI/ML and related areas of computing. This also underscores the value of incorporating multiple perspectives when auditing syllabi. 

Figure~\ref{fig:strength_and_gaps} summarizes the average normalized scores for each criterion (0–1 scale) across the evaluators. The use of an \emph{Institutional Template} and strong \emph{Department Affiliation} for the courses received consistently high scores, reflecting standardized practices. However, weaknesses emerge in \emph{Clarity of Course Goals} and recognition of \emph{Instructor’s Status (Tenure)}. The lack of explicit learning objectives risks obscuring how justice principles are meant to be applied in technical contexts, leaving students without clear guidance on how to contextualize justice-oriented practices and use ethical considerations. Similarly, the absence of transparency about an instructor’s position (\emph{Instructor Status (Tenure)}) may inadvertently conceal the power structures shaping whose perspectives are legitimized in the classroom. These underrepresented practices suggest opportunities for improvement, particularly in making justice-oriented objectives explicit and situating them within the positionality of those teaching the course.

\subsubsection{RQ2: What justice-oriented practices can be consistently identified across syllabi of courses meant to explicate ideas of fairness and ethics in AI/ML and related areas of computing, and which remain underrepresented?}

\begin{figure}[!ht]
\includegraphics[width=0.4\textwidth]{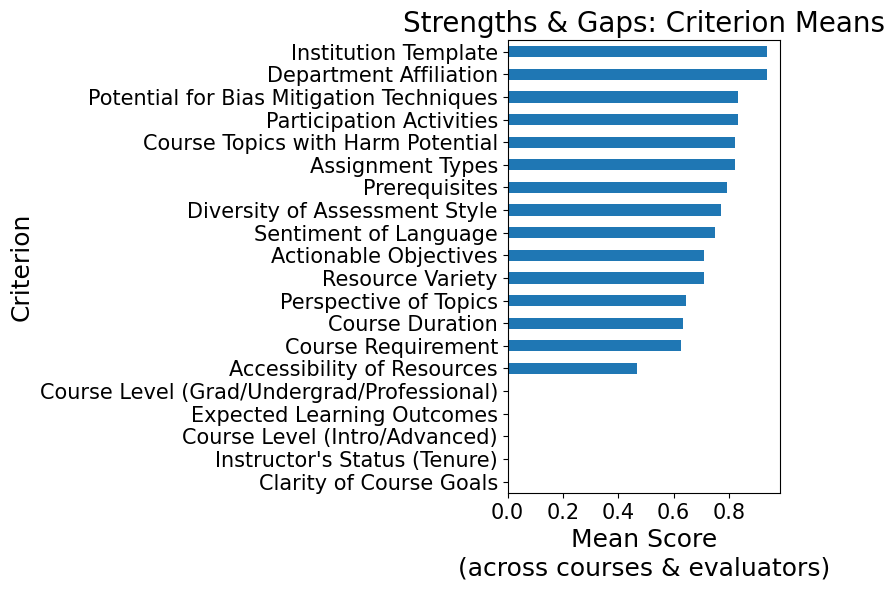}
\caption{Mean syllabus scores by criterion (0–1 scale) across evaluators and courses. }
\label{fig:strength_and_gaps}
\end{figure}

Referring again to Figure~\ref{fig:strength_and_gaps}, several justice-oriented practices were consistently identified across the syllabi. These include the use of neutral, non-biased language (\emph{Sentiment of Language}), acknowledgement of \emph{Bias Mitigation Techniques} and \emph{Harm Potential}, and high levels of \emph{Diversity of Assessment Style}, \emph{Resource Variety}, and \emph{Assignment Types}. Their consistent presence suggests that instructors recognize the importance of exposing students to multiple perspectives, diverse resources, and varied modes of evaluation in fairness and ethics courses in AI/ML and related areas of computing. This may increase the likelihood that students encounter justice-related issues from different vantage points, which is an important consideration in such courses. This has no bearing on the level at which these items are taught; however, they are at least acknowledged as attributes which should be discussed during such courses.

\subsubsection{RQ3: How are fairness, bias, and justice positioned in fairness and ethics in AI/ML and related computing courses, and how do they explicate these ideas?}
\begin{figure}[!ht]
    \centering
    \includegraphics[width=0.4\textwidth]{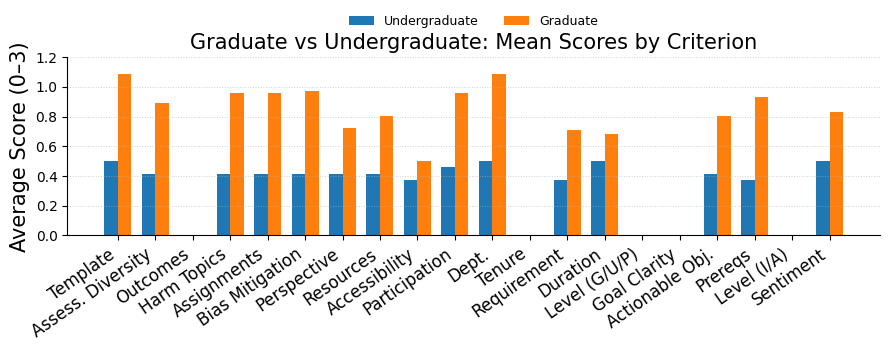}
    \caption{Graduate vs. Undergraduate syllabi mean scores (0–3) per criterion, across evaluators. }
    \label{fig:grad_under}
\end{figure}

Figure~\ref{fig:grad_under} compares mean criterion scores between graduate and undergraduate syllabi, averaged across evaluators. Graduate courses generally scored higher across most rubric dimensions, especially in having engaging/varied \emph{Participation Activities} and clearly denoting the course's specific \emph{Department Affiliation}, suggesting highly intentional design in course topics and clearer expectations. Undergraduate courses scored relatively low across the board, with close similarities to graduate courses in \emph{Accessibility of Resources} and having diverse \emph{Perspective of Topics}. This suggests foundational knowledge of fairness principles is being introduced to undergraduate students; these critical aspects that truly capture the essence of fairness and ethics are often absent, leading to an incomplete foundation that hampers a full and meaningful understanding of the core principles.
\begin{figure*}[th!]
\centering
\begin{minipage}{0.23\textwidth}
    \centering
    \includegraphics[width=.95\linewidth]{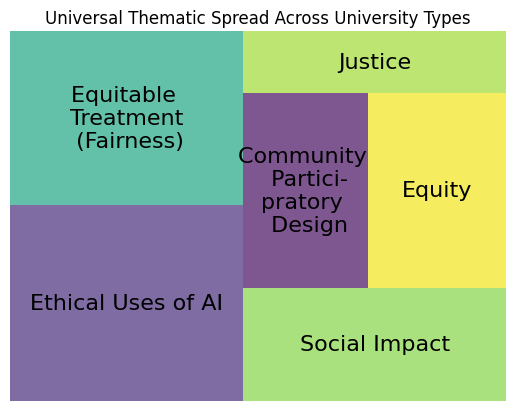}
\caption{Thematic Trends Across All Syllabi Focused on Courses in Fairness \& Ethics, in AI/ML and related areas of computing}
    \label{fig:universaltheme}
\end{minipage}\hfill
\begin{minipage}{0.23\textwidth}
    \centering
    \includegraphics[width=.95\linewidth]{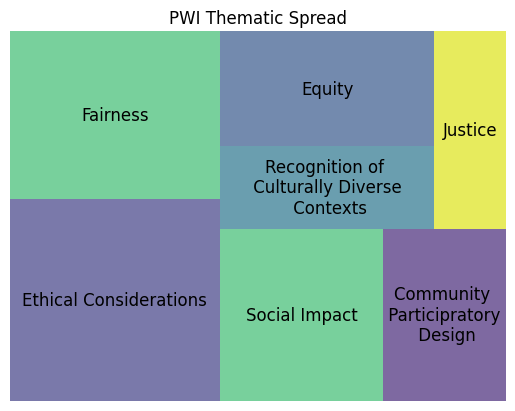}
    \caption{Thematic Trends Across Syllabi from PWIs Focused on Courses in Fairness \& Ethics, in AI/ML and related areas of computing}
    \label{fig:pwi_theme}
\end{minipage}\hfill
\begin{minipage}{0.23\textwidth}
    \centering
    \includegraphics[width=.95\linewidth]{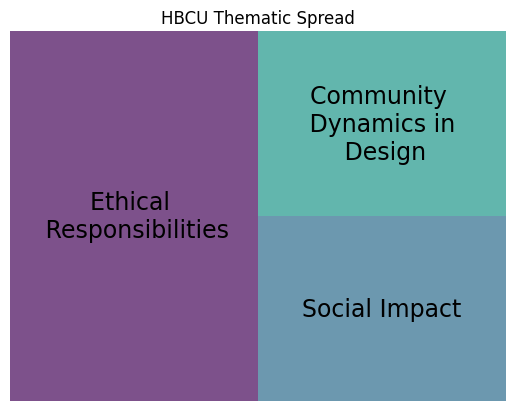}
    \caption{Thematic Trends Across Syllabi from an HBCU Focused on Courses in Fairness \& Ethics, in AI/ML and related areas of computing}
    \label{fig:hbcu_theme}
\end{minipage}\hfill
\begin{minipage}{0.22\textwidth}
    \centering
    \includegraphics[width=\linewidth]{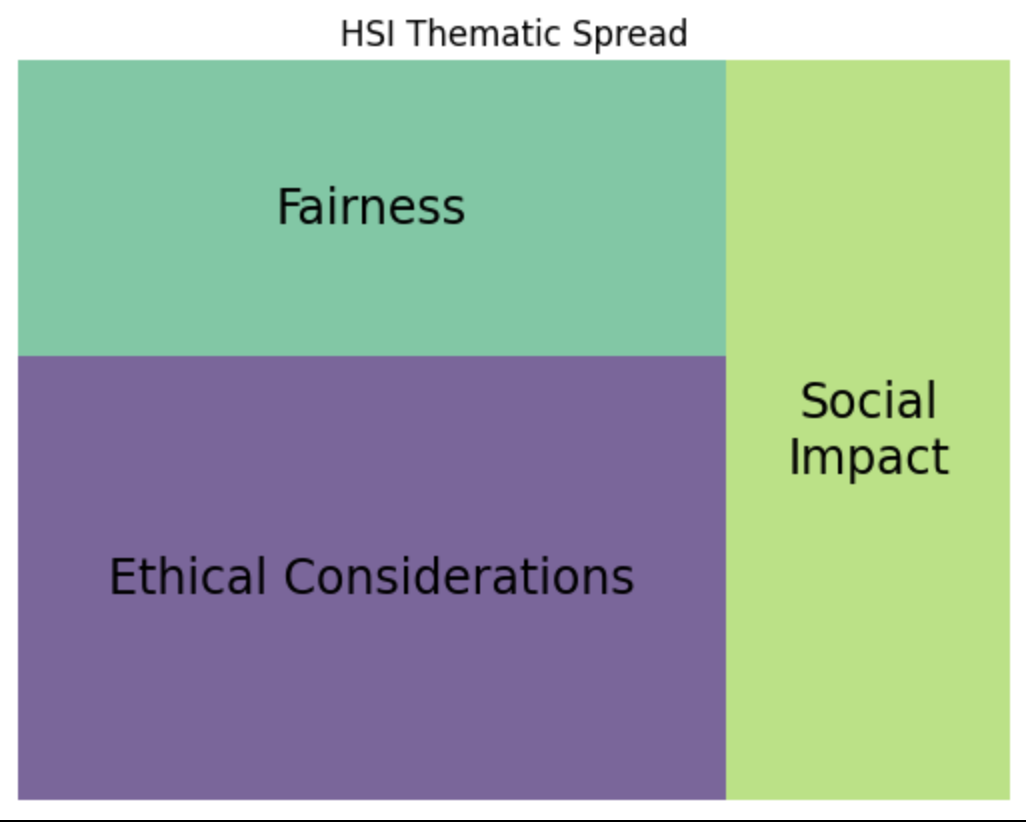}
    \caption{Thematic Trends Across Syllabi from HSIs Focused on Courses in Fairness \& Ethics, in AI/ML and related areas of computing}
    \label{fig:hsi_theme}
\end{minipage}
\end{figure*}

\subsection{Thematic Output Analysis}
\subsubsection{RQ2: What justice-oriented practices can be consistently identified across syllabi of courses meant to explicate ideas of fairness and ethics in AI/ML and related areas of computing, and which remain underrepresented?}

\paragraph{Representation.}
The most common main course readings across the courses we analyzed were ``Weapons of Math Destruction" by Cathy O'Neil (4 courses), ``Fairness and Machine Learning" by Barocas, Hardt, and Narayanan (3 courses), and ``Trustworthy Machine Learning" by Kush Varshney (2 courses). We note that one of the three these assigned readings is authored by a scholar from an underrepresented background in mathematics and computing. Others of marginalized backgrounds are included as supplemental or weekly readings, but not as the main text for the course. Including their voices in fairness and ethics curricula is especially important, as their perspectives offer unique insights into the challenges and nuances of fairness. Indeed, the very need for these conversations often stems from the kinds of inequities and harms they are likely to have personally experienced or witnessed. Thus, their voices add necessary, valuable insight to setting the tone of the courses we analyzed. We also note that at least one of the most commonly used resources, \emph{Trustworthy Machine Learning}, references many works by a diverse range of voices and includes a panel of diverse voices to ensure diversity of thought throughout the work, an important form of representation. The LLM also suggested that several resources are unique to individual courses, which explains why so few courses share resources.

\subsubsection{RQ3: How are fairness, bias, and justice positioned in fairness and ethics in AI/ML and related computing courses, and how do they explicate these ideas?}
The most frequently occurring justice-oriented terms across our syllabi included `learning,' `fairness,' `social,' `ethics,' `project,' and `impact.' These words are visualized in a word cloud shown in Figure~\ref{fig:wordcloud} of the Appendix, along with additional common words present in the syllabi, where the size of each word corresponds to its frequency. Note that `justice' is not a term that is widely used when discussing fairness and ethics. 
\paragraph{Institutional Trends.}
When conducting university-type specific analysis, the LLM found the most popular topics to be `ethics', `fairness', and `social impact'. Figure~\ref{fig:universaltheme} shows how these and other themes are represented across all of the courses we analyzed. We discuss how each institution type (i.e., PWI, HSI, HBCU) engages with these topics throughout this section. The least engaged topics were `justice' and `deconstructing colonial structures'. The LLM noted that when justice is mentioned as a framework for the courses, it involves the legal and moral implications of technology, ensuring accountability and protection of rights and when discussing deconstructing colonial structures few syllabi explicitly mention the need to challenge colonial ideologies and structures in the context of fairness and ethics in technology.

Upon drilling down to trends in the syllabi we analyzed from PWIs, the LLM found the most represented syllabi were ones that discussed `fairness' and `ethical uses of AI', whilst `justice' was the least explored. The LLM denoted most of the discussions on fairness surrounded commonly used definitions, measures, and mitigation strategies for equitable treatment; discussions on ethical uses of AI considered `algorithmic bias', `fairness', and `privacy'; and four institutions mentioned justice, focusing on AI and data science with legal implications.

Across the HSI institutions in our sample, the LLM found three popular theme types: `ethics', `fairness', and `social impact'. The LLM denoted most of the discussions on ethics focused on ethical considerations in data science and AI, particularly in relation to marginalized groups; fairness discussions focused on addressing algorithmic fairness and bias in data-driven decision-making; and the focus on social impact was directed by the application of ethical frameworks to enhance social good through technology. Note in Figure~\ref{fig:hsi_theme} that fairness and social impact are represented equally. We also note that HSIs serve a population that adopts practices of social cohesion~\cite{ruiz2021exploring, almeida2009multilevel}. The topics covered in this syllabus appear to encourage active participation and empowerment with the use of AI/ML frameworks taught in the classroom.

When analyzing our syllabi from an HBCU, `ethics' with a particular emphasis on ethical responsibilities in computing, particularly in relation to marginalized communities; `social impact' with application of computing ethics to enhance social justice and equity; and `community engagement' eloquating the importance of community dynamics in technology development were the most common topics. The representations of these topics in this syllabus align with some of the goals of many HBCUs: education as a means of addressing inequalities and engaging in community activism~\cite{crewe2017education}. They also often encourage a sense of responsibility for one's community~\cite{franklin2023role}. 

\paragraph{Undergraduate and Graduate Course Experiences.}
When comparing the experiences of fairness and ethics courses between undergraduate and graduate courses, the LLM uncovered $11$ out of the $18$ graduate courses had justice elements, while $3$ out of the $6$ undergraduate courses had elements of justice. For graduate courses, the focus was largely on fairness (10 courses), ethics (8 courses), transparency (5 courses), and accountability (4 courses). Other interesting themes that appeared in our analysis were social justice, privacy, bias mitigation, explainability, governance, algorithmic fairness, algorithmic bias, data mining, machine learning, robustness, interdisciplinary discussions, trustworthiness, ethical approaches, responsible data science, critical perspectives, public interest technology, civic data, professional responsibility, and power. For undergraduate courses, the focus was largely on ethics (2 courses) and social impact (2 courses). Other interesting themes were civic data, power, fairness, and professional responsibility. Fairness was discussed less at the undergraduate level (33\%) and more at the graduate level (55\%), while ethics was discussed less at the graduate level (44\%) and more at the undergraduate level (50\%). Civic data, power, and professional responsibility were equally discussed topics across the course levels.

\section{Discussions \& Recommendations}
\label{sec:discussion}
Our multi-perspective, LLM-assisted syllabus analysis shows that evaluation outcomes depend as much on who is evaluating as on the content itself. Weak-to-moderate agreement across evaluators highlights the influence of institutional familiarity and positional context, raising questions about reproducibility in AI-supported educational review. We also observed syllabi often lack clear goals and accessibility measures and Instructors and Internal Evaluators identify subtle justice-oriented practices more readily than Chairs or External evaluators. Graduate courses outperform undergraduate ones across nearly all criteria, revealing systemic differences in formality and comprehensiveness which may affect student preparedness to engage in AI/ML discussions around fairness and ethics. Based on our findings, we offer several recommendations for syllabi used in fairness and ethics courses in AI, ML, and algorithmic design: (1) include evaluators from marginalized backgrounds and across institutional and external levels, (2) encourage instructors to use explicit, justice-oriented language and practices (like the ones we discuss) in course objectives, (3) apply a justice-oriented rubric (like ours) to assess and refine syllabi, and (4) integrate explicit discussions of justice, which are often missing but provide critical perspectives in such syllabi.

\section{Limitations \& Future Work}

While our approach leverages LLMs to simulate evaluator perspectives, these simulations are influenced by prompt design and inherent model biases. Human oversight remains important for refining prompts and interpreting ambiguous LLM outputs not addressed in this study. Future work could involve larger teams to analyze bigger datasets, extend this framework to disciplines beyond AI/ML, incorporate more personas (e.g., student perspectives), and assess how multi-perspective feedback influences syllabus design.

\section{Conclusion}
\label{sec:conclusion}
This study demonstrates how LLM-assisted evaluation can complement human review in assessing course syllabi across multiple perspectives and course themes by systematically comparing evaluator agreement, identifying syllabus strengths and gaps in justice-oriented criteria of AI, ML, and algorithmic design for fairness and ethics courses. Our findings highlight the potential of LLMs to scale diagnostic reviews while revealing the importance of human calibration, institutional context, and justice-aware criteria for such courses.

\bibliography{reference}

\label{sec:reference_examples}

\appendix

\onecolumn

\section{Appendix} \label{app:process_score_rubric}

\subsection{Justice Scoring Rubric}
In Tables~\ref{tab:rubric1} and \ref{tab:rubric2}, we present the detailed scoring rubric used in our analysis. The rubric was developed through two iterative rounds involving all members of our research team, which includes individuals from diverse backgrounds such as education and computer science. In the first iteration, team members collaboratively drafted an initial rubric. Each member then manually reviewed it by applying it to a small set of course syllabi. The goal was to gather feedback on how the rubric could be improved for clarity, coverage, and ease of application. Following several discussions and refinements, we finalized a second version of the rubric. This revised rubric was again tested by the team through an initial evaluation of different course syllabi. After confirming that the criteria and descriptions were clear and comprehensive, we adopted this final rubric for our study. Using carefully designed prompts that represented different evaluator roles, we instructed the LLM to assess each syllabus by assuming a specified persona and applying the rubric criteria accordingly. We include the rubric in full here to ensure transparency, enable replication of our process, and provide a reusable resource for future justice-oriented syllabus evaluations.

\begin{table*}[ht!]
\centering
\caption{Scoring rubric for syllabus evaluation (Part 1). Each criterion is scored from 0 (Not Evident) to 3 (Fully Evident).}
\label{tab:rubric1}
\resizebox{\textwidth}{!}{
\begin{tabular}{p{5cm} p{3cm} p{3.5cm} p{3cm} p{3cm}}
\toprule
\textbf{Criteria} & \textbf{0 – Not Evident} & \textbf{1 – Partially Evident} & \textbf{2 – Evident} & \textbf{3 – Fully Evident} \\
\midrule
\textbf{Institution Template} \\ Does the institution have a template for structuring syllabi? & No template provided / Unable to gauge source & Limited guidance, but no formal template & Template suggested but not required & Clear institutional template provided and required \\
\midrule
\textbf{Diversity of Assessment Style} \\ Does the course use traditional and non-traditional assessment methods 
(e.g., Traditional: exams, quizzes, essays | Non-Traditional: peer assessment, journal, debate)? & Only traditional assessments used & Majorly traditional assessments used, few non-traditional & Majorly non-traditional assessments used, few traditional & Diverse range of non-traditional and traditional assessments \\
\midrule
\textbf{Expected Learning Outcomes (specific, measurable aims)} \\ Are the Learning Outcomes connected to algorithmic fairness? & No clear connection with algorithmic fairness & Vague or general outcomes connected to algorithmic fairness & Clear outcomes connected to algorithmic fairness & Detailed, competency-specific outcomes connected to algorithmic fairness \\
\midrule
\textbf{Course Topics with Harm Potential} \\  Are there topics that could promote harm or disparity, and are these harms discussed?
(e.g., demographic parity, equal opportunity, etc...) & No awareness of potential harm or disparities & Topics may introduce harm, with limited discussion & Harmful/disparate topics acknowledged but minimally addressed & Harmful topics identified and fully discussed with mitigation strategies \\
\midrule
\textbf{Assignment Types} \\ Are there variety in assignments are used (e.g., hands-on, presentations)?& No assignments or all one type & Limited assignment variety & Multiple types of assignments, some variety & Diverse assignments (e.g., hands-on, presentations, group work) \\
\midrule
\textbf{Potential for Bias Mitigation Techniques} \\  How is bias mitigation being introduced or tackled in the course content? & No consideration of potential bias & Limited awareness, minimal mitigation & Some awareness of bias, partially addressed & Full awareness of bias potential, with mitigation strategies \\
\midrule
\textbf{Perspective of Topics} \\ From what perspective are topics introduced? Who is telling the story/being promoted?
Who is the knowledge holder/being promoted in the syllabus? & All sources have similar demographics, biased perspective & Limited diversity in perspectives, biased toward one viewpoint & Multiple perspectives with some inclusivity & Broad, inclusive perspectives actively promoted \\
\midrule
\textbf{Resource Variety} \\ What/How many resources are provided for each topic? & No resources provided & Minimal/limited types of resources & Some variety in resources (e.g., scholarly work from 1–2 populations) & Extensive, varied resources (articles, videos, case studies) \\
\bottomrule
\end{tabular}
}
\end{table*}

\begin{table*}[ht!]
\centering
\caption{Scoring rubric for syllabus evaluation (Part 2).}
\label{tab:rubric2}
\resizebox{\textwidth}{!}{
\begin{tabular}{p{5cm} p{3cm} p{3.5cm} p{3cm} p{3cm}}
\toprule
\textbf{Criteria} & \textbf{0 – Not Evident} & \textbf{1 – Partially Evident} & \textbf{2 – Evident} & \textbf{3 – Fully Evident} \\
\midrule
\textbf{Accessibility of Resources} \\ Are there accessibility considerations in resources (e.g., language, hearing, sight)? & No accessibility considerations & Minimal accessibility adjustments & Accessibility addressed in some areas (e.g., captions, alt text) & Full accessibility support (language, vision, hearing adaptations) \\
\midrule
\textbf{Participation Activities} \\ What kinds of activities are contributing to participation, and in what ways? & No participatory activities & Minimal/optional participation activities & Some participatory activities, limited engagement & Engaging, varied participation activities (e.g., discussions, group work) \\
\midrule
\textbf{Department Affiliation} \\ What department offers the course? & Department not indicated & Department mentioned but unclear & Department stated, relevance implied & Clearly stated department affiliation and relevance \\
\midrule
\textbf{Instructor’s Status (Tenure)} \\ Is the instructor tenured or non-tenured? & No indication of tenure status & Few instructor statuses indicated & Some instructor statuses indicated & Tenure status explicitly stated for all instructors \\
\midrule
\textbf{Course Requirement} \\ Is this an elective or a required course? & Not specified if elective or required & Mentioned but unclear & Clearly mentioned as elective or required & Clear and emphasized as elective/required with curriculum relevance \\
\midrule
\textbf{Course Duration} \\ What is the duration of the course? & Duration not specified & Duration vaguely indicated & Clear duration but not emphasized & Duration fully stated and relevant to structure \\
\midrule
\textbf{Course Level (Grad/Undergrad/Professional)} \\ Is this course for grad, undergrad, or professional students? & Course level not indicated & Course level implied but unclear & Clearly stated level & Explicitly stated mix of levels \\
\midrule
\textbf{Clarity of Course Goals} \\ How clear are the goals of the course? & Goals unclear or absent & Goals vaguely defined & Goals clear but general & Specific, well-defined, measurable goals \\
\midrule
\textbf{Actionable Objectives} \\ Are there objectives that discuss actionability/applicability of the work? & No mention of applicability/action & Some objectives suggest application & Clear objectives, some application focus & Objectives emphasize applicability and actionability \\
\midrule
\textbf{Prerequisites} \\ Are there prerequisites for this course? & No prerequisites mentioned & Prerequisites vaguely referenced & Clear prerequisites but not detailed & Prerequisites clearly stated \\
\midrule
\textbf{Course Level (Intro/Advanced)} \\ Is this an introductory or advanced topic? & Level not stated & Level implied but unclear & Clearly stated level & Explicitly stated and relevant to structure \\
\midrule
\textbf{Sentiment of Language} \\ What is the sentiment of the wording? & No sentiment analysis & Limited sentiment awareness & Sentiment generally neutral & Positive or constructive sentiment throughout \\
\bottomrule
\end{tabular}
}
\end{table*}

\clearpage
\subsection{Persona Prompting}
We show all promptings used to simulate each persona to the LLM.

\begin{tcolorbox}[colback=gray!10!white, colframe=black, title=LLM Prompt: Internal Evaluator, sharp corners=southwest, boxrule=0.5pt]
You are an \emph{expert-level educational evaluator internal to the university} of the AI/ML syllabus that you belong to, which may be inside or outside of your department. Your task is to assess this university course syllabus using the rubric provided. Your goal is to assess if the course is promoting diverse ideas from scholars of multiple backgrounds, inclusive pedagogy,decolonized approaches, accommodate different learning needs, and create space for critical engagement and safe learning. For each criterion, assign a score from 0 (not evident) to 3 (fully evident), and provide a short, specific justification for your ranking.

\end{tcolorbox}

\begin{tcolorbox}[
    colback=gray!10!white,
    colframe=black,
    title=LLM Prompt: External Evaluator,
    sharp corners=southwest,
    boxrule=0.5pt,
    left=2pt,          %
    right=2pt,         %
    top=2pt,           %
    bottom=2pt         %
]
You are an \emph{expert-level external educational evaluator for the university} from which this AI/ML syllabus is coming from. Your task is to assess this university course syllabus using the rubric provided. Your goal is to assess if the course is for the sake of justice-oriented practices such as promoting diverse ideas from scholars of multiple backgrounds, inclusive pedagogy,decolonized approaches, accommodate different learning needs, create space for critical engagement, and a safe learning environment. For each criterion, assign a score from 0 (not evident) to 3 (fully evident), and provide a short, specific justification grounded in the content of the syllabus for your ranking.
\end{tcolorbox}

\begin{tcolorbox}[colback=gray!10!white, colframe=black, title=LLM Prompt: Department Chair, sharp corners=southwest, boxrule=0.5pt]
You are a \emph{departmental chair} internal to the university of the AI/ML syllabus that you belong to, which is inside of your department. Your task is to assess this university course syllabus using the rubric provided. Your goal is to assess if the course is promoting diverse ideas from scholars of multiple backgrounds, inclusive pedagogy,decolonized approaches, accommodate different learning needs, and create space for critical engagement and safe learning. For each criterion, assign a score from 0 (not evident) to 3 (fully evident), and provide a short, specific justification for your ranking.

\end{tcolorbox}

\begin{tcolorbox}[colback=gray!10!white, colframe=black, title=LLM Prompt: Instructor, sharp corners=southwest, boxrule=0.5pt]
You are an experienced \emph{course instructor}, you could be a professor or lecturer of a different course inside of the department. Your task is to assess this university course AI/ML syllabus using the rubric provided. Your goal is to assess if the course is promoting diverse ideas from scholars of multiple backgrounds, inclusive pedagogy, decolonized approaches, accommodate different learning needs, and create space for critical engagement and safe learning. For each criterion, assign a score from 0 (not evident) to 3 (fully evident), and provide a short, specific justification for your ranking.

\end{tcolorbox}

\subsection{Theme Prompting}
We first explored if there were any general trends and statistically significant findings we could gather from our set of syllabi with a simple prompt as shown here in \textbf{LLM Thematic Prompt: Qualitative Finding}.

\begin{tcolorbox}[colback=gray!10!white, colframe=black, title=LLM Thematic Prompt: Qualitative Findings, sharp corners=southwest, boxrule=0.5pt]

Give me important statistics across these syllabi {json\_string}. Quantify with numbers and statistics. Tell me if anything is statistically significant.
\label{box:themegen}
\end{tcolorbox}

\begin{center}
\begin{tcolorbox}
[
    colback=gray!10!white,
    colframe=black,
    title=LLM Thematic Prompt: Quantitative Findings,
    sharp corners=southwest,
    width=\linewidth,  %
    boxrule=0.1pt,
    left=1pt,          %
    right=1pt,         %
    top=1pt,           %
    bottom=1pt         %
]
\small
You are an expert syllabus evaluator. Find themes in these sets of syllabi: [json\_string]. Compare the themes across each university type, county by type (e.g., cultural, community engagement, equity, fairness, ethics, social impact, etc.). Explicitly quantify as many themes as possible for reporting, do not list the course name, do list the university and actual counts (do not say many, several, etc.). Do not summarize, but report. Your focus in reporting is the university type and themes. You at least need to check if the material comes from an approach that uplifts and acknowledges the efforts of marginalized people in these spaces. You can do this by checking if they prioritise centring the perspectives, knowledge, and experiences of historically marginalized groups and actively working to deconstruct and resist colonial structures and ideologies in fairness (so you will need to analyze the authors of the resources for the class). Also, check what common terms are used across universities and departments by types [(MSI, HBCU, PWI, 4-year, Ivy League, 2-year, Community college)/(R1, R2, course-level, race/gender of instructor)], quantify and explain what the terms are by count and by university name and type. Report if the course focuses on equality (concepts that force equality, such as demographic parity, equal opportunity), equity (concepts that allow for thresholding of fairness, such as intersectional fairness, disparate impact, equalized odds), or justice (concepts that allow for justice).
\end{tcolorbox} 
\end{center}

\clearpage

\subsection{Additional Thematic Analysis}
\begin{figure}[h]
    \centering
    \includegraphics[width=0.75\textwidth]{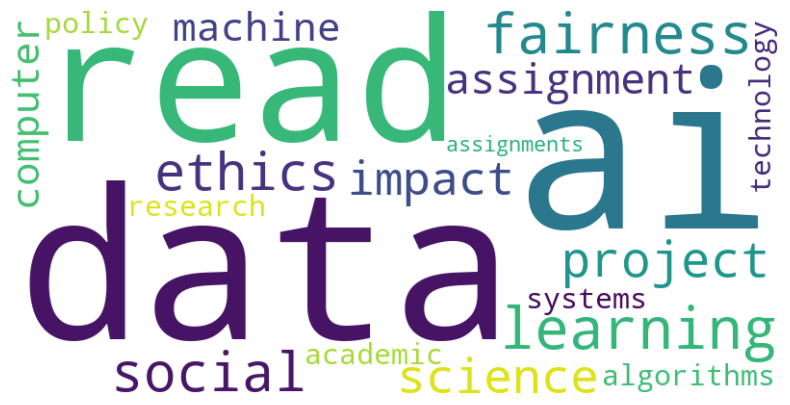}
    \caption{Wordcloud with the most common words in our set of syllabi, with the size of each word correlating to the frequency of the word's appearance.}
    \label{fig:wordcloud}
\end{figure}

There are several words which are common across the syllabi we study. In Figure~\ref{fig:wordcloud}, we show the most common words in our set of syllabi, with the size of each word correlating to the frequency of the word's appearance.

\end{document}